\documentclass[12pt]{article}
\usepackage{epsf}
\usepackage{graphicx}
\evensidemargin - 0.8 cm
\newcommand{\be}{\begin{equation}}
\newcommand{\ee}{\end{equation}}
\newcommand{\bea}{\begin{eqnarray}}
\newcommand{\eea}{\end{eqnarray}}
\newcommand{\ba}{\begin{array}{ccc}}
\newcommand{\ea}{\end{array}}
\newcommand{\nn}{\nonumber}

\newcommand{\eq}[1]{eq.~(\ref{#1})}
\newcommand{\eqs}[2]{eqs.(\ref{#1}, \ref{#2})}

\newcommand{\fig}[1]{Fig.~(\ref{#1})}

\def\log{\textnormal{log}}
\def\det{\textnormal{det}}
\def\exp{\textnormal{exp}}
\def\ok{\omega_k}
\def\ge{\gamma_{{\rm E}}}

\def\Tr{ {\rm Tr} }
\def\Sp{{\rm Sp}}

\def\cos{{\rm cos}}

\def\sin{{\rm sin}}
\def\A{{\mathcal A}}

\def\npb#1#2#3{Nucl. Phys. B {\bf #1}, #2 (#3)}

\def\pr#1#2#3{Phys. Rev. {\bf #1}, #2 (#3)}

\def\prd#1#2#3{Phys. Rev. D {\bf #1}, #2 (#3)}

\def\rmp#1#2#3{Rev. Mod. Phys.{\bf #1}, #2 (#3)}

\begin{document}
%%%%%%%%%%%%%%%%%%%%%%%%%%%%%%%%%%%%%%%%%%%%%%%

\topmargin -1.4cm
\oddsidemargin -0.8cm
\evensidemargin -0.8cm
\title{\Large{{\bf The kinetic energy for the static SU(2) Polyakov line}
\\ 
}}

\vspace{1.5cm}

\author{~
{\sc Michaela Oswald\footnote{Work done in collaboration with
Dmitri Diakonov, NORDITA, diakonov@nordita.dk}}\\
{\small Niels Bohr Institute, Blegdamsvej 17, 2100 Copenhagen, Denmark}}

\date{} 
\maketitle
\vfill
%%%%%%%%%%%%%%%%%%%%%%%%%%%%%%%%%%%%%%%%%%%%%%%%
\begin{abstract} 
\noindent
At very high temperatures Yang--Mills theories can be described through perturbation theory. At the tree level the time components of the gluon fields decouple and yield a dimensionally reduced theory. The expectation value of the Polyakov loop then assumes values of the $Z(N)$ center group. At intermediate temperatures, however, this is not true anymore. The time dependence shows up in loops. In a recent work \cite{DO} we integrated out fast varying quantum fluctuations around background $A_i$ and static $A_4$ fields. We assumed that these fields are slowly varying but that the amplitude of $A_4$ is arbitrary. As a result we obtained the kinetic energy terms for the Polyakov loop both for the electric and the magnetic sector of SU(2). 
\end{abstract}

%%%%%%%%%%%%%%%%%%%%%%%%%%%%%%%%%%%%%%
\vfill

\thispagestyle{empty}
\newpage

%%%%%%%%%%%%%%%%%%%%%%%%%%%%%%%%%%%%%%
\section{Introduction}
%%%%%%%%%%%%%%%%%%%%%%%%%%%%%%%%%%%%%%

The partition function of a Yang--Mills theory at finite temperature in its Euclidean-invariant form is given by
\bea\label{pf}
{\cal Z}=\int DA_\mu\,\exp\left\{-\frac{1}{4g^2}\int_0^{\beta=\frac{1}{T}}\!\!\!\!\!dt \int d^3x\,F_{\mu\nu}^aF_{\mu\nu}^a\right\}.
\eea
The gluon fields obey periodic boundary conditions in the temporal direction. In this form the theory is usually simulated on the lattice. Yang--Mills theory is of particular interest since it has a phase transition from a confined to a deconfined phase. The order parameter is the so-called Polyakov loop, which is defined as the trace of the Polyakov line
\bea
P(x)={\rm P}\,\exp\left(i\int_0^{1/T}\!dt\,A_4\right).
\eea
At very high temperatures the potential energy of the Polyakov line (or of $A_4$) has its zero-energy minima for values of $P(x)$ at the center of the gauge group (or for quantized values of $A_4$). High temperature perturbation theory hence corresponds to the system oscillating around these trivial values of the Polyakov line, {\it i.e.} $<\Tr\,P \neq 0>$. As the temperature decreases, however, the fluctuations of the Polyakov line increase and eventually at the critical temperature $T_c$ the system undergoes a phase transition from a deconfined to a confined phase which has $<\Tr\,P = 0>$. In order to approach this phase transition from the high-temperature side, one needs to study the Polyakov line in its whole range of possible variation.  In our recent paper \cite{DO} we worked with static and diagonal $A_4$ gluon fields and the gauge group $SU(2)$. In this case the Polyakov line
\be
P(x)=\exp\left(i\frac{A_4(x)}{T}\right)
\ee
has the gauge invariant eigenvalues 
\bea\label{eigenvalues}
e^{\pm i\pi\nu}\qquad {\rm where}\qquad \nu=\sqrt{A_4^aA_4^a}/2\pi T .
\eea
We assume that the gluons are varying slowly, but we allow for an arbitrary amplitude of the $A_4$ fields. We then find the non-trivial effective action for the eigenvalues of the Polyakov line, interacting in a covariant way with the spatial gluon fields $A_i$.

%%%%%%%%%%%%%%%%%%%%
\section{The 1-loop action}
%%%%%%%%%%%%%%%%%%%%

Since the gluon fields in the partition function \eq{pf} are periodic fields they can be decomposed into Fourier modes: 
\be
A_\mu(t,x)=\sum_{k=-\infty}^\infty A(\ok,x)\,e^{i\ok t},
\qquad {\ok=2\pi k\,T,}
\ee
where the $\ok$ are the so-called Matsubara frequencies, which play the
role of mass. 
The first step on the way to an effective theory is to integrate out the non-zero Matsubara modes. This reduces the original 4D Euclidean symmetry to a 3D one. This procedure is exact in the $T\to\infty$ limit. At intermediate temperatures, however, the nonzero modes show up in loops and produce infinitely many effective vertices. In our work \cite{DO} we found all these vertices, but restricted to low momenta $p<T$. 

%%%%%%%%%%%%%%%%%%%%%%%%%
\subsection{The background field method}
%%%%%%%%%%%%%%%%%%%%%%%%%

In order to calculate the quantum fluctuations we decompose all the gluon fields into background fields (denoted by a bar) and quantum fluctuations around them:
\be
A_\mu =\bar A_\mu + a_\mu . 
\ee
The fluctuations are not gauge-invariant and we choose the background Lorentz gauge $D_\mu^{ab}(\bar A)\,a_\mu^b=0$, where
\be
D_\mu^{ab}(\bar{A}) = \partial_{\mu}\delta^{ab} + f^{acb}\bar{A}_{\mu}^{c}\;.
\ee
is the covariant derivative in the adjoint representation. 

A one loop calculation corresponds to expanding the partition function to quadratic order in the fluctuations $a_\mu$. Doing that one obtains
 \be\label{pfbf}
Z(\bar{A}) = e^{S} = e^{\bar{S}}\, \int{}Da\,D\chi\,D\chi^{+}\,\exp \left\{-
\frac{1}{2g^2(M)}\int{}d^4 x\,a_{\mu}^b\, W_{\mu\nu}^{bc}\, a_{\nu}^c - \int \, d^4
x\,\chi^{+a}\left(D^2_{\mu}\right)\,\chi^{a}\right\}\;,
\ee
where $\chi, \chi^+$ are ghost fields and 
\be
 \bar{S} =  -\frac{1}{4 g^2(M)}\int\,d^4 x{}F_{\mu\nu}^{a}(\bar{A}) F_{\mu\nu}^{a}(\bar{A}) 
\ee
is the action of the background fields. The quadratic form for $a_{\mu}$ is given by
\be\label{W}
W_{\mu\nu}^{ab} = -[ D^2(\bar{A})]^{ab} \delta_{\mu\nu} - 2 f^{acb}F_{\mu\nu}^c(\bar{A}) \;.
\ee
For the 1-loop action we have to integrate out the quantum fluctuations of the gluons and the ghost degrees of freedom. This results in
\be\label{S1loop}
S_{\rm{ 1-loop}} = \log\, \left(\det{W}\right)^{-1/2} + \log\,\det \left(-D^2\right)\;.
\ee
Since the only gluon fields which are left are the background fields we will omit the bar from now on. 

So far the background fields have been kept arbitrary.  One can, however, always choose the gauge where $A_4(x)$ is static.  The spatial components are in principal time dependent, although periodic in the time direction, since any time-independent gauge transformation will generate a time dependent $A_i(t,x)$. 

In our calculation, however, it turns out to be easy to reconstruct the results for a time dependent $A_i(t,x)$ from an original ansatz where the spatial components are static as well. We hence assume static background fields in what follows, and will give the more general results at the end. 

%%%%%%%%%%%%%%%%%%%%%
\section{Gradient expansion of $S_{{\rm 1-loop}}$}
%%%%%%%%%%%%%%%%%%%%%
The effective action for the eigenvalues of the Polyakov line is given in terms of a potential term and a kinetic one, where the latter contains electric and magnetic field strengths. In order to obtain these terms we expand the 1-loop action \eq{S1loop} in powers of $D_i$ using that:
\be\label{El-Mag}
E_i = F_{i4} = D_i A_4 \qquad {\rm and}\qquad B_k^a = \frac{1}{2}\epsilon_{ijk} F_{ij}^{a}=\frac{1}{4}\epsilon_{ijk}\epsilon^{cad}[D_i,D_j]^{cd}.
\ee
Since for the case of $SU(2)$ only two independent color vectors exist in the electric (magnetic) sector, namely $E_i$ ($B_i$) and $A_4$ we expect that the gradient expansion takes the following form:
\bea\label{gradex}
S_{\rm{ 1-loop}}\!\! =\! \int\frac{d^3 x}{T} \left[-T^4\,V(\nu)\! +\!
  E_i^2 f_1(\nu)\! +\! \frac{(E_i A_4)^2}{A_4^2} f_2(\nu)\! +\! B_i^2\,h_1(\nu)\! +\!
\frac{(B_i A_4)^2}{A_4^2}\,h_2(\nu)\! +\! \ldots\! \right].
\eea
Here $V(\nu)$ is the potential energy of the eigenvalues of the Polyakov line, \eq{eigenvalues}, which are given in term of the rescaled gluon field variable $\nu=\sqrt{A_4^aA_4^a}/2\pi T$. The potential term has been known previously \cite{GPY, Weiss}, and the functions $f_{1,2}, h_{1,2}$ are the new findings of \cite{DO}.

%%%%%%%%%%%%%%%%%%%%%
\subsection{Schwinger's proper time formalism}
%%%%%%%%%%%%%%%%%%%%%

In order to attack the logarithm of the functional determinants in \eq{S1loop} we avail ourselves of a method that was originally introduced by Schwinger \cite{Schw}. In addition we want to both normalize and regularize \eq{S1loop}, i.e. we subtract the free zero-gluon contribution and introduce a Pauli-Villars cutoff $M$. For the ghost contribution this results in
\bea\label{ghost}
\lefteqn{\log\,\det (-D^2)_{\rm{Norm, Reg}} \equiv \log\,
\frac{\det (-D_{\mu}^2)}{\det (-\partial_{\mu}^2)}\,\frac{\det
 (-\partial_{\mu}^2 + M^2)}{\det (-D_{\mu}^2 + M^2)}}\\ 
&& 
= 
-\int_0^{\infty}\frac{ds}{s}\;  \Sp \left[\underbrace{\left( 1 - e^{-s
        M^2}\right)}_{{\rm Pauli-Villars}} \,\left( e^{s D_{\mu}^2} 
- \underbrace{e^{s \partial_{\mu}^2}}_{{\rm free}} \right)\right]\;.
\eea
Here $M$ denotes the Pauli-Villars mass, and with Sp we denote a functional trace. To evaluate the trace we insert a plane wave basis and find:
\bea
\lefteqn{\log\,\det (-D^2)_{\rm{Norm, Reg}} = -\int d^3 x \sum_{k=-\infty}^{\infty}\int\frac{d^3 p}{2\pi^3}\int_0^{\infty}\frac{ds}{s}\left( 1 - e^{-s M^2}\right) }\\ \nn
&&
\times\Tr \left\{\exp\left[s (\A^2 + (D_i+i p_i)^2) \right]- \exp\left[-s (\ok^2 + p^2)\right]\right\}, 
\eea
where we defined the adjoint matrix $\A^{ab} = f^{acb} A_4^c + i \omega_k \delta^{ab}$. 
Similarly we obtain for the gluon determinant:
\bea\label{gluon}
\lefteqn{\log\,(\det W)^{-1/2}_{{\rm Norm, Reg}} = 
\frac{1}{2}\int\,d^3 x\,\sum_{k=-\infty}^{\infty}\int\frac{d^3 p}{(2
 \pi)^3}\,\int_0^{\infty}\frac{ds}{s}\left( 1 - e^{-s M^2}\right)}
\\&\times&
\Tr\,\left\{\exp\left[s\left( (\A^2 + (D_i+i p_i)^2)^{ab} \delta_{\mu\nu} +
2f^{acb}F_{\mu\nu}^c\right)\right]
-  \exp\left[-s(\ok^2+p^2)\right]\right\}\;.\nn
\eea
So far expressions (\ref{ghost}, \ref{gluon}) are independent of the gauge group, but it should be noticed that all matrices are in the adjoint representation. 

%%%%%%%%%%%%%%%%%%%%%%%%%%%%
\subsection{Effective potential}
\begin{figure}[]
\centerline{
\epsfxsize=0.35\textwidth
\epsfbox{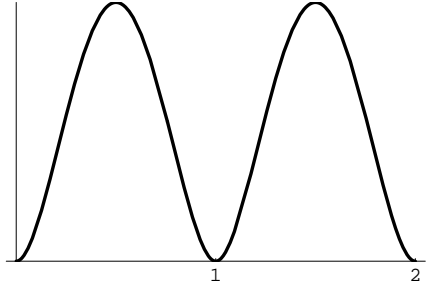}}
\caption{ {\small The periodic potential $V$ with period 1 in units of $\nu$}.\label{Potential} }
\end{figure}

%%%%%%%%%%%%%%%%%%%%%%%%%%%%
In the zero-gradient order one has to set $D_i=0$. One finds 
$\bf E=\bf B=0$ and 
\be
\det^{-\frac{1}{2}}W_{\mu\nu} = \det^{-2}(-D_{\mu}^2).
\ee
For the explicit calculation one can choose a gauge where $A_4$ is diagonal in the fundamental representation, which for the gauge group $SU(2)$ means
\be
A_4^a = \delta^{a3}\phi = \delta^{a3}2\pi T \,\nu\qquad, \qquad \nu=\frac{\sqrt{A_4^a A_4^a}}{2\pi T}. 
\ee
The resulting potential is well know \cite{GPY, Weiss} and reads
\be
V = \frac{1}{3(2\pi)T^4}\;\phi^2(2\pi T-|\phi|)^2|_{\rm mod\;2\pi T} = \frac{(2\pi)^2}{3}\,\nu^2 (1- \nu)^2|_{\rm mod\,1}.
\ee 
This result is plotted in \fig{Potential}. It is clearly periodic in $\nu$ with period one. This reflects the center symmetry of the theory. 
At the minima of the potential $A_4$ has quantized values which corresponds to a Polyakov line with values at the center of the group, which is $Z(N)$ for $SU(N)$. The group $Z(N)$ is a discrete one and has the elements $z_k=\exp(2\pi i k/N)$, where $k=\{0,\ldots, N-1\}$. For $SU(2)$ this means that
\bea
P=\exp\left(iA_4^a\frac{\tau^a}{2T}\right)=\cos\frac{|A_4|}{2T}+i\frac{A_4^a\tau^a}{|A_4|}
\sin\frac{|A_4|}{2T},
\eea
\bea
|A_4|&=&0,4\pi T,\ldots : P=\left(\begin{array}{cc}1&0\\0&1\end{array}\right),\\
|A_4|&=&2\pi T,6\pi T,\ldots : P=\left(\begin{array}{cc}-1&0\\0&-1\end{array}\right).  
\eea

%%%%%%%%%%%%%%%%%%%%%%%%%%%%%%%%
%%%%%%%%%%%%%%%%%%%%%%%%%%%%%%%%

%%%%%%%%%%%%%%%%%%%%%%%%%%%%
\subsection{Second derivatives}
%%%%%%%%%%%%%%%%%%%%%%%%%%%%
When we go to higher orders of the covariant derivative the calculation becomes more elaborate. We want to keep all powers of $A_4$ but expand in $D_i$. The {\it technique}  is to expand e.g. 
\bea
\exp\,s\!\left({\A}^2\!+\!(D_i\!+\!ip_i)^2\right),\qquad
{\A}^{ab}&=&f^{acb} A_4^c + i \omega_k \delta^{ab}
\eea
in powers of $D_i$ using 
\bea\label{master1}
e^{A+B}&=&e^A+\!\int_0^1\!d\alpha\, e^{\alpha A}\,B\,e^{(1-\alpha)A}  \\ \nonumber
&+& \!\int_0^1\!d\alpha\!\int_0^{1-\alpha}\!d\beta\, e^{\alpha A}\,B\,
e^{\beta A}\,B\,e^{(1-\alpha-\beta)A}+\ldots ,
\eea
and drag $B=D_i, D_i^2$ to the right using
\bea\label{master2}
[B,e^A]=\int_0^1\!d\gamma\,e^{\gamma A}\,[B,A]\,e^{(1-\gamma)A} .
\eea
Then we have to evaluate all the integrals over $\alpha, \beta, \gamma, \ldots,  p, s$ and sum over the Matsubara frequencies $\ok = 2 \pi k T$. This should be done separately for the ghost and gluon determinants. 

%%%%%%%%%%%%%%%%%%%%%%%%%%%%
\section{Results for the electric sector}
%%%%%%%%%%%%%%%%%%%%%%%%%%%%
In the expansion of \eqs{ghost}{gluon}  by means of \eqs{master1}{master2}  we identify the electric field \eq{El-Mag} in the following structures:
\bea\label{el1}
[D_i,{\A}]=[D_i,D_4]=-iF_{i4}=-iE_i ,
\eea
\bea\label{el2}
[D_i, \A^2]= - i \left\{ \A, E_i\right\}\, ,
\eea
where all matrices are in the adjoint representation, e.g.  $E_i^{ab}= i f^{acb} E_i^c$.  At this point we can make the generalization to a time dependent background $A_i(t,x)$ field. Since $\A^{ab} = D_4^{ab}+ i \omega_k \delta^{ab}$ the time dependence shows up in the covariant time derivative $D_4^{ab} = \partial_4  \delta^{ab} + f^{acb}A_4^c$. The functional form of \eqs{el1}{el2} remains unaltered but  from now on all our results will be valid for a general time dependent electric field:
\be\label{etime}
E_i^a=D_i^{ab}A_4^b-\dot A_i^a=\partial_iA_4^a+\epsilon^{acb}A_i^cA_4^b-\dot A_i^a.
\ee

After all integrations and the summation over $\ok$, both for $\log\,\det\,(-D^2)$ and $\log(\det\,W)^{1/2}$ the coefficients for the electric sector (see \eq{gradex}) are found to be \cite{DO}
\bea
f_1 &=& \frac{11}{48 \pi^2}\left[2 \left( \log\,\mu - \ge\right) - \psi\left(-\frac{\nu}{2}\right) -
\psi\left(\frac{\nu}{2}\right) + \frac{20}{11\nu}\right] \label{f1}\;, \\
f_2 &=& \frac{11}{48 \pi^2}\left[\psi\left(-\frac{\nu}{2}\right) +
  \psi\left(\frac{\nu}{2}\right) -\psi\left(\nu\right) -
  \psi\left(1-\nu\right) - \frac{20}{11\nu}\right] \label{f2}.
\eea
Here $\psi$ is the digamma function,
\be
\psi(z) = \frac{\partial}{\partial\,z}\log\,\Gamma(z)\;,
\ee
$\ge$ is the Euler constant and $\mu$ is a UV cutoff that we introduced in the sum over Matsubara frequencies. 
It is related to the Pauli-Villars mass as
\be\label{mu}
\mu= \frac{M}{4\pi T}\,e^{\ge}\,.
\ee
This scale has been previously found in \cite{coupling} for the running coupling constant in the dimensionally reduced theory, and our result agrees.

\begin{figure}[t]
\centerline{
\epsfxsize=0.45\textwidth
\epsfbox{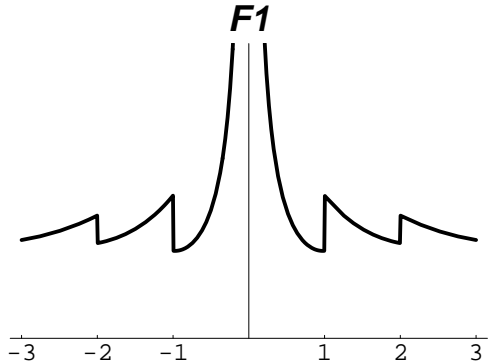}}
\caption{ {\small The function $f_1(\nu)$ with the constant part subtracted, in different
intervals.}\label{F1interval} }
\end{figure}

It should be noted here that the above results are valid for $0\leq \phi\leq 2\pi T$, i.e. $0\leq\nu\leq 1$. In other intervals the functional forms of $f_1$ and $f_2$ are different, it is the sum over the Matsubara frequencies that causes this. We show the results for $f_1$  for a broader range of $\nu$ in \fig{F1interval}.
%%%%%%%%%%%%%%%%%%%%%%%%%%%%%%%%
%%%%%%%%%%%%%%%%%%%%%%%%%%%%%%%%
One can clearly see that this result is not $Z(2)$ -symmetric, the same is true for $f_2$.  However, one particular combination, namely 
\be
f_3\equiv f_1 + f_2 = \frac{11}{48 \pi^2}\left[2 \left( \log\,\mu - \ge\right) -
  \psi\left(\nu\right) - \psi\left(1-\nu\right)\right]
\ee
turns out to periodic.  We plot it in \fig{SymmDiff}. The reason for this is the following: 
We chose the gauge for the $A_4$ fields where they are static and diagonal in the fundamental representation.  This leaves certain residual gauge symmetries left, namely
\be\label{resid}
A_\mu\to S^\dagger A_\mu S+iS^\dagger\partial_\mu S,\qquad 
S(x,t)=\exp\left\{-i\frac{\tau^3}{2}\left[\alpha(x)+2\pi tTn\right]\right\}.
\ee
Our invariants in the electric sector can be expressed as
\be
E_i^a E_i^a f_1 + \frac{(E_i^a A_4^a)^2}{A_4^b A_4^b} f_2 = E_i^\parallel E_i^\parallel f_3 + E_i^\perp E_i^\perp f_1
\ee
where $E_i^\parallel E_i^\parallel = (E_i^1)^2+  (E_i^2)^2$ and $E_i^\perp E_i^\perp=(E_i^3)^2$ denote the structures parallel and orthogonal to $A_4^2$. 

The time-dependent gauge transformations \eq{resid} now introduce large time derivatives in the $A_i^{1,2}$ but not in the $A_i^3$ fields.  The time-dependent part of the electric field \eq{etime} enters in $E_i^\perp E_i^\perp$, but not in $E_i^\parallel E_i^\parallel$.  Hence one should not expect gauge-invariance in the structure $E_i^\perp E_i^\perp f_1$, since it is only quadratic in $\dot A_i^{1,2}$. In order to have gauge invariance one would have to sum over all powers $\dot A_i^{1,2}/T$, which would result in a non-local effective action. 

%%%%%%%%%%%%%%%%%%%%%%%%%%%%
\section{Results for the magnetic sector}
\begin{figure}[t]
\centerline{
\epsfxsize=0.45\textwidth
\epsfbox{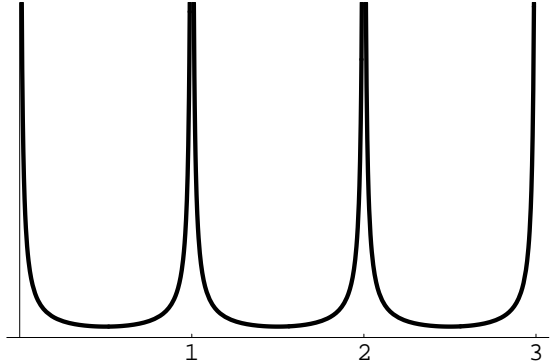}}
\caption{ {\small The symmetric function in $f_3$, $h_{1,2}$ without the constant part, in different intervals.
}\label{SymmDiff} }
\end{figure}

%%%%%%%%%%%%%%%%%%%%%%%%%%%%
Since the magnetic field (\ref{El-Mag}) does not contain any explicit time-dependence we expect the functions $h_{1,2}$ to be periodic in $\nu$. Indeed this turns out to be the case. The functional form of $h_{1,2}$ depends again on the interval that we choose for $A_4$. For $0\leq\nu\leq 1$ we obtain
\bea\label{h1}
h_1(\nu) &=& \frac{11}{96
\pi^2}\left[4\left(\log\,\frac{M}{4 \pi T} + \frac{\ge}{2}\right) - \psi\left(\nu\right) -
  \psi\left(1-\nu\right) \right] \,, \\
\label{h2}
h_2(\nu) &=& -
\frac{11}{96 \pi^2}\left[2\ge + \psi\left(\nu\right) +
  \psi\left(1-\nu\right)\right]\,.
\eea
In \fig{SymmDiff} we plot the constant part (which is the same as for $f_3$) for different intervals. The result is obviously center-symmetric. 
%%%%%%%%%%%%%%%%%%%%%%%%%%%%%%%%
%%%%%%%%%%%%%%%%%%%%%%%%%%%%%%%%

%%%%%%%%%%%%%%%%%%%%%%%%%%%%
\section{Renormalization}
%%%%%%%%%%%%%%%%%%%%%%%%%%%%
In the coefficients that multiply $E_i^2$ and $B_i^2$ (\eqs{f1}{h1}) we had to introduce a UV cutoff $\mu$, see \eq{mu}. This 1-loop divergence is necessary to cancel the tree level divergence, which comes from the running of the running coupling constant once we quantize the theory. The point is to set the scale of the coupling constant equal to the Pauli-Villars mass $M$:
\bea
-\frac{F_{\mu\nu}^a\,F_{\mu\nu}^a}{4 g^2(M)} = -F_{\mu\nu}^a\,F_{\mu\nu}^a\, \frac{11}{3}{\rm N_c}\frac{1}{32 \pi^2}\,\log\frac{M}{\Lambda}\;.
\eea
If we add the tree level and the 1-loop results we find for the constant parts containing the logarithm in the kinetic energy:
\bea
\frac{11}{24\pi^2 T}\,\log\frac{\Lambda}{4\pi T} \left(E_i^a E_i^a + B_i^a B_i^a\right) 
\eea
which is definitely finite.
%%%%%%%%%%%%%%%%%%%%%%%%%%%%
\section{Comparison to previous work}
%%%%%%%%%%%%%%%%%%%%%%%%%%%%
In reference \cite{Chapman} the author makes a covariant derivative expansion of the 1-loop Yang-Mills action. While we keep all powers of the background $A_4$ field the author of \cite{Chapman} goes only to quadratic order. For a comparison we have to expand our functions $f_{1,2}$ and $h_{1,2}$ to quadratic order in $\nu$. The results agree exactly with \cite{Chapman}, for details see \cite{DO}.  

We mentioned in the section on the electric sector that one combination of our functions, namely $f_3=f_1+f_2$ is $Z(2)$ symmetric. This function has been obtained in \cite{BGK-AP} in the context of a calculation of the interface tension of $Z(N)$ instantons, and our result again agrees.

%%%%%%%%%%%%%%%%%%%%%%%%%%%%
\section{Summary}
%%%%%%%%%%%%%%%%%%%%%%%%%%%%

In our recent paper \cite{DO} we studied the effective action for the eigenvalues of a static $SU(2)$ Polyakov line at high temperatures. In the $T\to\infty$ limit perturbation theory works and dimensional reduction takes place. The Polyakov loop has values at the center of the gauge group. If one lowers the temperature, however, the fluctuations of the Polyakov loop around these perturbative values increase. We studied the fluctuations of $P$ around its perturbative values and found the 1-loop effective action for its eigenvalues, interacting with the $A_i$ fields. We find that the the kinetic energy in the electric sector is not center-symmetric. If one wishes to preserve this symmetry one has to sum over all powers of the electric field, which results in a non-local effective theory. 
 
\vspace{1cm}
%%%%%%%%%%%%%%%%%%%%%%%%%%%%%%%%%%%%%%%%
{\sc Acknowledgments:}\\
I would like to thank the organizers for inviting me to Zakopane, and I am grateful to my supervisor Dmitri Diakonov for providing me with this interesting problem and for innumerable helpful and useful discussions.  

%%%%%%%%%%%%%%%%%%%%%%%%%%%%%%%%%%%%%%%%
%%%%%%%%%%%%%%%%%%%%%%%%%%%%%%%%

\end{document}